\documentclass[lettersize,journal]{IEEEtran}
\usepackage{amsmath,amsfonts}
\usepackage{algorithmic}
\usepackage{algorithm}
\usepackage{array}
\usepackage[caption=false,font=normalsize,labelfont=sf,textfont=sf]{subfig}
\usepackage{textcomp}
\usepackage{stfloats}
\usepackage{multirow}
\usepackage{url}
\usepackage{verbatim}
\usepackage{graphicx}
\usepackage{cite}
\usepackage{etoolbox} 
\makeatletter 
\patchcmd{\@makecaption} 
  {\scshape} 
  {} 
  {} 
  {} 
\makeatother 
\usepackage{fancyhdr} 
\usepackage{hyperref}
\hypersetup{
    colorlinks=true,
    linkcolor=blue,
    filecolor=green,
    citecolor=green
}
\fancypagestyle{FirstPage}{
\lfoot{978-1-6654-8453-4/22/\$31.00 ©2022 IEEE} 
}
\begin{document}

\title{LBDMIDS: LSTM Based Deep Learning Model for Intrusion Detection Systems for IoT Networks}

\author{\author{Kumar Saurabh$^1$, Saksham Sood$^1$, P. Aditya Kumar$^1$, Uphar Singh$^1$, Ranjana Vyas$^1$, O.P. Vyas$^1$, Rahamatullah Khondoker$^2$ \\ $^1$Indian Institute of Information Technology, Allahabad, India\\$^2$Department of Business Informatics, THM University of Applied Sciences, Friedberg, Germany\\ pwc2017001@iiita.ac.in, iit2019164@iiita.ac.in, iit2019165@iiita.ac.in, pse2017003@iiita.ac.in, ranjana@iiita.ac.in, opvyas@iiita.ac.in, rahamatullah.khondoker@mnd.thm.de}}


\maketitle

\begin{abstract}
%
 In the recent years, we have witnessed a huge growth in the number of Internet of Things (IoT) and edge devices being used in our everyday activities. This demands the security of these devices from cyber attacks to be improved to protect its users. For years, Machine Learning (ML) techniques have been used to develop Network Intrusion Detection Systems (NIDS) with the aim of increasing their reliability/robustness. Among the earlier ML techniques DT performed well. In the recent years, Deep Learning (DL) techniques have been used in an attempt to build more reliable systems. In this paper, a Deep Learning enabled Long Short Term Memory (LSTM) Autoencoder and a 13-feature Deep Neural Network (DNN) models were developed which performed a lot better in terms of accuracy on UNSW-NB15 and Bot-IoT datsets. Hence we proposed LBDMIDS, where we developed NIDS models based on variants of LSTMs namely, stacked LSTM and bidirectional LSTM and validated their performance on the UNSW\_NB15 and BoT\-IoT datasets. This paper concludes that these variants in LBDMIDS outperform classic ML techniques and perform similarly to the DNN models that have been suggested in the past.
\end{abstract}

\thispagestyle{FirstPage}
\vspace{-0.3cm}
\section{Introduction}

Internet of Things (IoT) is a collection of devices which gather and share data over the Internet. Data is gathered using sensors, which are embedded in these devices. IoT devices are used in many areas, ranging from household devices like smart watches, smart bulbs, smart air conditioners, temperature sensors to more complex devices like smart vehicles and smart electrical grids. They are also extensively used in manufacturing, transportation, infrastructure, military equipment and healthcare. It was estimated that there are over 35 billions IoT devices would be in use upto 2021 \hyperlink{ref1}{[1]}. This high number has contributed to increasing number of cyber attacks on IoT networks, which demands the security of these devices from cyber attacks to be increased as the current security measures have proven to be inadequate \hyperlink{ref2}{[2]}.
\par Network Intrusion Detection Systems (NIDS) are used to detect cyber attacks and malicious activities like Denial of Service (DoS), Distributed DoS (DDoS), Worms, Backdoor, etc. Network traffic is monitored and potential threats are identified. Signature-based NIDS is good at detecting known attacks but fail at detecting attacks which have not been seen before. Anomaly-based NIDS is good at detecting new attacks as they focus more on attack patterns but give high false positives.

 \par Therefore, NIDS integrated with ML and DL techniques have been developed to better identify new threats \hyperlink{ref3}{[3]}. Earlier, ML and DL methods have been applied unanimously to develop systems capable of detecting intrusion in the networks for conventional and extensive communication. However, every method came with limitations of its own and gradually the need for a better system increased as the classical models failed in terms of handling the heterogeneity of data. Also, the dataset used here should have multiple types of attack vectors (like DoS, DDoS, Worms) to tackle multi-classification of attack categories. Traditional models are unable to detect zero-day attacks as the dataset is unable to hold the entries which are new to the attack analysis.
\par So, a smart model is needed which can detect any anomaly which rises from any deviation from normal behavior of the associated network. So, this hypotheses could also detect zero-day attack as the model will not solely depend on the pre-built classes of attacks. However, this will again give rise to a problem which will bias towards the general class. This will lead to raised false positive rates. To get an intermediate model which can act more sustainable, DL needs to be implemented.
\par Classic ML techniques have been used in this field for 20+ years while keeping the KDD99 dataset in consideration. But, with the onset of technological boom, the attack categories are increasing spontaneously. So, the potential of any classic ML model is far smaller than the reach of the intrusion. So, methods like SVM (Support Vector Machine), DT (Decision Tree), KNN (K-Nearest Neighbor), etc. have been ruled out long before when considering for industrial applications. Also, the hybrid method using these models has worked well for a long time, but eventually falls short in front of recent DL models. The basis for working of an ML model is basically supervised and unsupervised learning techniques other than the reinforcement learning part. ML extensively depends upon how rich the dataset is available to us. Also, the inability to scale the data accordingly also poses a large limitation to the extensive use of any ML model.
\par To handle such issues, DL methods like ANN (Artificial Neural Networks) came into picture. Here the model is built on neurons which are coordinated by parameters and hyper-parameters. To scale the input and use it on extensive scale, the number of layers are kept accordingly to get maximum efficiency. DL methods have proved far better than the ML models in terms of accuracy, precision, etc. with the ability to handle large amounts of data. ANN, CNN (Convolutional Neural Networks), RNN (Recurrent Neural Networks), FDN (Feed Forward Deep Networks) are many examples of DL architectures. DL techniques in general have outperformed ML techniques.
\par As research progressed, it was seen that basic DL architectures lacked the ability to detect unknown attacks and even if they did, the false positives and false negatives were very high. To tackle this problem, LSTM (Long Short-Term Memory) was introduced. This methodology is a special form of RNN. The most distinguishing feature of LSTM is to keep the information/parameter for later use in the system. Thus, they can handle the data which is time series and could be variable with respect to time.
\par The primary motivation of using LSTM lies in its approach where is is not restricted to the limitations of conventional DL (neural networks) methods. Here, the input sequences and output sequences between layers are variable which could accordingly work efficiently to detect known as well as unknown attacks.
\par The paper is organized as follows: Section II contains related research which demonstrates the earlier work done, Section III describes the datasets used in our experiments, Section IV includes our proposed methodology, Section V contains the results and performance analysis, and Section VI gives the conclusion and future works.
\vspace{-0.3cm}
\section{Literature Review}

\par In paper \hyperlink{ref4}{[4]}, IDS was developed to learn the behavior of normal network traffic. Dataset used was UNSW-NB 15 for communication of external networks. The methods being experimented here were Artificial Immune System (AIS), Filtered-based SVM (FSVM),  Euclidean distance Map (EDM), and Geometric Area Analysis (GAA) which performed 85\%, 92\%, 90\% and 93\% respectively. In \hyperlink{ref5}{[5]}, BoT-IoT and UNSW-NB15 datasets are considered in the experiment. After the data standardization is done, the MLP (Multi-Layered Perceptron) model is used followed by adapting the hyperparamters. The comparison was based on the BoT-IoT dataset while the classifier was built on UNSW-NB15 dataset.
ARM (Association Rule Mining) and Naïve Bayes only had 85\% and 72\% accuracy respectively. The normal perceptron model was basic and gave 63\% accuracy but when converted to the prima facie 13- feature DNN model, it gave 99\% accuracy. Among the ML techniques, Decision tree gave the highest accuracy (93\%). The process of training the models consumes a lot of time and memory \hyperlink{ref5}{[5]}. In paper \hyperlink{ref6}{[6]}, the authors explore ML based approach. On carrying out an experiment it was seen that the ML methods along with flow identifiers were effective in detecting botnet attacks. Four ML algorithms were used namely, ARM, Artificial Neural Network (ANN), Naïve Bayes and Decision Tree. Decision Tree gave the best results with the highest accuracy of 93.23\% and the lowest False Alarm Rate of 6.77\% whereas ANN was the least accurate with an accuracy of 63.97\% and False Alarm Rates of 36.03\%.
In \hyperlink{ref7}{[7]}, the authors have addressed the issue of lack of a data set which can appropriately show the modern  network traffic and attacks. So this paper looks into the creation of a UNSW-NB15 dataset. It has 49 features developed using Argus and Bro-IDS tools.
In paper \hyperlink{ref8}{[8]}, the authors proposed a hybrid IDS for IoT networks by combining CNNs (Convolutional Neural Network) and LSTM. They used UNSW\_NB-15 dataset and compared the performances of the hybrid IDS and RNN for binary classification, and the accuracy achieves was 95.7\% and 98.7\% respectively.
In \hyperlink{ref9}{[9]}, the authors compared the performances of DL methods like RNN, GRU and Text-CNN with some of the traditional ML methods on the KD99 and ADFA-LD datasets. The DL methods were able to outperform the traditional ML techniques. In paper \hyperlink{ref10}{[10]}, the authors implemented Linear Discriminant Analysis (LDA),
Classification and Regression Tree (CART) and Random Forest (RF) on KD99 dataset. The accuracy achieved were 98.1\%, 98\% and 99.65\% respectively.
In \hyperlink{ref11}{[11]}, the authors implemented a Feed-Forward Neural Network for binary and multi-class classification on the Bot-IoT dataset for normal and three different attack categories (DDoS/DoS, Reconnaissance, Information Theft). They were able to achieve 98\%, 99.4\%, 98.4\% and 88.9\% accuracy on the normal and the three attack categories respectively.
In paper \hyperlink{ref12}{[12]}, the authors proposed an IDS based on blockchain and deep learning. The DNN model was able to achieve 98\% accuracy for binary classification and 97\% accuracy for multi-class classification on the NSL-KDD dataset.


\section{Dataset Explanation}
In order to train our models and to determine their reliability in testing phase, Data selection is necessary. In the past datasets like KDD98, KDDCUP99 and NSLKDD were used as comprehensive datasets for Network Intrusion Detection System(NIDS). In the recent times the researches have shown that the these datasets don't reflect the modern network traffic (normal and attack vectors). Hence, we selected a few datasets that were made publicly available by researchers in the last few years. These datasets contained labelled network data that were generated in labs with the help of virtual network setup. The datasets are a hybrid of Normal network traffic and synthetic botnet attack traffic. The datasets selected are: 
\subsection{UNSW-NB15 Dataset}
Widely in use since 2015 as a dataset for evaluating NIDS, it was created by Cyber Range Lab of UNSW-Canberra \hyperlink{ref14}{[14-17]}. The Dataset contains 49 (excluding class labels) features and a total 2,540,044 records that were split across 4 CSV files. The dataset contains a total of 10 class labels out of which there are 9 types of Botnet attack vectors and 1 Normal class. The different types of attack traffic are as follows: analysis, fuzzers, backdoor, generic, Denial of Service (DoS), reconnaissance, shellcode, exploits and worms \hyperlink{ref13}{[13]}. Out of the 49 features 13 important ones were selected in the paper \hyperlink{ref4}{[4]}, namely source ip address, source port, destination ip address, destination port, duration, source bytes, destination bytes, source TTL, destination TTL, Source load, Destination load, Source packets and Destination packets.
\vspace{-0.3cm}
\subsection{Bot-Iot Dataset} 
Created by the Cyber Range Lab of UNSW-Canberra in 2019, it includes network and botnet traffic incorporated into a network environment. The Dataset contains 47 features including class label, attack category and attack subcategory \hyperlink{ref18}{[18]}. A total of 72 million records of Normal and attack traffic constitute the Dataset. A pre-selected training and testing dataset which includes 10-best features was created which have been used for training and validation purposes. The training and testing dataset have a combined of 3.6 million records. The Dataset includes attack vectors of types DDoS, DoS, OS and service scan, keylogging and Data exfiltration attacks. On the basis of protocol used DDoS and Dos categories are further divided into different types.
\begin{figure*}[!t]
\centering
\includegraphics[width=7in]{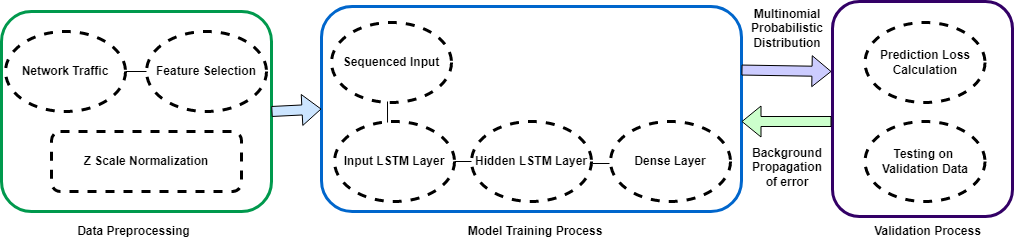}
\label{fig_first_case}
\begin{center}
\caption{Proposed LBDMIDS architecture}
\end{center}
\label{fig_sim}
\end{figure*}
\section{Proposed Methodology}

\subsection{Working Principles}
Let us understand the working principles behind our project. We are using two variants of LSTM in our research, Stacked and Bidirectional. First, let us see Feed Forward Neural Networks and Recurrent Neural Networks.

\subsubsection{Feed Forward Neural Networks}
This model consists of input layer and output layer with several hidden layers in between. This model is generally designed to contain three or more hidden layers. The layers consist of biases and weights. The input data is injected into the first layer and consequently the output is generated which becomes the input for the next layer. The number of nodes are adjusted so that they match the number of features in input data. The weights and biases used in the layers are initialised randomly but later optimised through a method known as back-propagation.

Important note here is that the model is “one-way” and there are no backward connections here. The data computation flows in one direction where any input generates an output where there is no connection such that the output could be injected back to the model. The absence of backward connections give rise to limitations such as the "error-sum" problem where the error generated in one layer keeps on increasing in the consequent layers. As there is no option of taking the feedback from the output, the input and output become independent which makes the tuning of hyper-parameters by the model an un-achievable task.

\subsubsection {Recurrent Neural Networks}
RNN is a type of network normally distinguished by “memory”. The layers present in the model usually take account of the previous inputs while considering the input of a certain layer. While traditional neural networks work on the principle that input and output are generated independently, RNN works on the method of feeding the output back into the system which generally works in favor of reducing the error and balancing the model.

Another variation which makes RNN different from FFNN (Feed Forward Neural Networks) is the weights and parameters associated with the layers of network. Each node in FFNN uses different weights which are adjusted through the processes of gradient descent and back-propagation. In RNN, the parameters are shared between the layers of network, although the parameters are still adjusted by the same processes used in FFNN (gradient descent and back-propagation). 

While taking the “feedbacks” from the succeeding layers, RNN is able to adjust its parameters by analysing the errors encountered by the model. This works well in maintaining the error percentage to a threshold and also fits the model better.

Overcoming the problem faced by FFNN, RNN faces two other limitations. These are known as vanishing gradient and exploding gradient problem. These problems generally depend upon the function of error in the model. Here, gradient is defined as the slope of the error function. If the slope becomes too low, then it will continue to decrease as the threshold keeps decreasing. This will give rise to non-learning of the model. The parameters in the model will become negligible and the memory of model will saturate.
On the other hand, when slope becomes too high, the model will become unstable and the data fed to the model will generate fluctuating outputs which will in turn render the model parameters too large to be considered. 
\subsubsection{Long Short-Term Memory}

LSTMs are a special variant of RNN. These are particularly used in learning long-term dependencies. RNNs fail in learning information when the dataset is large and the entries have much similarity. So, LSTM was designed to handle information for longer feeds of input and work for larger epochs.

\par
As the layers of RNN have a very simple structure, for example a single Tanh layer \hyperlink{ref19}{[19]}, remembering information for a long period of time becomes a struggling task for them. On the other hand, in a general LSTM model, there are four interacting layers which work in a special way to remember the important information for a long duration of time.

\par
The problem of vanishing gradient faced by RNN is also solved by LSTM as they continue to learn new information while keeping the previous ones. Hence the significance of parameters is maintained and the model becomes stable. 

\par
LSTMs also have different variations. Although the difference is pretty small between those, they could be of great significance depending upon the input data fed to the network. Common types of LSTM are Classic, Stacked and Bidirectional. Stacked LSTM has several LSTM layers and can only access past samples. Bi-directional LSTM has two layers and has access to both past and future samples.
\vspace{-0.2cm}
\subsection{Network Intrusion Detection System(NIDS) Architecture}

We proposed NIDS based on two variants of LSTM - Stacked LSTM and Bi-directional LSTM to help filter out normal and attack vectors in IoT network traffic.
The workflow of the NIDS architecture can be divided into three phases, namely the data Preprocessing, Training of the Model and Model Validation as shown in Fig.1.
\vspace{-0.2cm}
\subsection{Data Preprocessing}
In Data Preprocessing stage, the raw data extracted from IoT networks in form of PCAP/CSV files is processed in a suitable form to be fed to the LSTM Model. The data is filtered to remove any redundancies and get rid of Null values. Afterwards the most important features are selected to be fed into the proposed model, which is then followed by z-score Normalization of feature which ensures similar distributions for each feature.
\subsubsection{Feature Extraction} We managed to reduce the Dimensionality of the raw data by selecting the most important features which made the data suitable for processing. Very often, large Datasets have a lot of redundant and correlated data that can be filtered out without losing important or relevant information. In case of USNW\_NB15, we selected the following most important 13 features  \hyperlink{ref1}{[1]}:
\begin{enumerate}
    \item{Source ip address - IP address of the attacker computer} 
    \item{Source port - Port number of the attacker computer}
    \item{Destination ip address - IP address of the victim computer}
    \item{Destination port - Port number of the victim computer} 
    \item{Duration - Record total duration of transaction} 
    \item{Source bytes - Number of bytes sent from source to the destination} 
    \item{Destination bytes - Number of bytes sent from destination to the source} 
    \item{Source TTL - Source to destination time to live value} 
    \item{Destination TTL - Destination to source time to live value} 
    \item{Source load - Transmission rate in bits per second}
    \item{Destination load - Reception rate in bits per second} 
    \item{Source packets - Number of packets sent from source to the destination} 
    \item{Destination packets - Number of packets sent from destination to the source}
\end{enumerate}{}{}

For Bot-IoT dataset, the 10 features that were pre-selected in the Reduced training and testing set were:
\begin{enumerate}
    \item {rate - Total packets per second in transaction}
    \item{srate - Source-to-destination packets per second}
    \item{drate - Destination-to-source packets per second}
    \item{min - Minimum duration of aggregated records}
    \item{max - Maximum duration of aggregated records}
    \item{mean - Average duration of aggregated records}
    \item{std\_dev - Standard deviation of aggregated records}
    \item{state\_number - Numerical representation of feature state}
    \item{flgs\_number - Numerical representation of feature flags}
    \item{seq - Argus sequence number}
\end{enumerate}{}{}

\subsubsection{z-scale Normalization}
The process of normalization is done to transform the data in a way they are distributed similarly. It helps the model to treat each feature with similar importance as it provides similar weights to each feature. Assuming the feature subspace has N rows and M columns i.e., X (feature subspace) = $\mathbb{R}^{N\times M}$, the z-scale normalization can be implemented as follows:
\noindent
\begin{align}
\mu_{m} &= \frac{\sum_{i = 0}^{N-1} x_{im}} {N} \\
\sigma_m &= \frac{\sum_{i=0}^{N-1}(x_{im} - \mu_{m})^{2}}{N} 
\end{align}
 
z-scale normalized feature vector can be obtained as \newline
\noindent
\begin{align}
z_{m} = \frac{X_{m}-\mu_{m}}{\sigma_m}
\end{align}
Here, $\mu_{m}$ is the mean of the entries of the m'th column and $\sigma_{m}$ is the  standard deviation of the entries of the m'th column.
\begin{algorithm}[H]
\caption{ Z-SCALE NORMALIZATION}\label{alg:alg1}
\begin{algorithmic}
\STATE 
\STATE {\textbf{for each col m in X(0,1,2...M-1)}}
\STATE \hspace{0.5cm}$ \mu_{m} \gets compute(1)$
\STATE \hspace{0.5cm}$ \sigma_{m} \gets compute(2)$
\STATE \hspace{0.5cm}$ z_{m} \gets compute(3)$
\STATE {\textbf{end for}}.
\end{algorithmic}
\label{alg1}
\end{algorithm}
\vspace{-0.2cm}
\subsection{Training and Validation Process}
Data Processing stage transforms the data into a more suitable form to be processed by the model which will lead to more accurate predictions. It is followed by changing the shape of the data to be processed by the LSTM layers. A suitable Timesteps parameter is selected and the dimensionality of the dataset is changed(samples, timesteps, features). Timesteps is the number of past samples on which the LSTM model looks back at.
The training and validation period of the model consists of feeding the time-series sequential data to the LSTM layers.
\subsubsection{Stacked LSTM} 
In case of stacked LSTM, there are multiple layers of LSTM stacked on top of one another. The LSTM layers help in uncovering the patterns and dependence of features to their class labels because they can learn at higher levels of abstractions. The input LSTM layer is followed by a batch of hidden LSTM layers that process sequenced input and combine learning patterns from the previous layers, to produce learning representations at higher levels of abstraction. The Dense layer is the final layer with the number of nodes equal to the number of categories in the output label. In the decision phase, the soft-max activation function\hyperlink {ref19}{[19]} is used by the dense layer to select the most probable of output classes, and the prediction error is calculated with the help of 'Sparse Categorical Crossentropy' which is then backpropogated to adjust the weights of the neural network. The model hyperparameters are shown in table 1.
\vspace{-0.3cm}
\begin{table}[ht!]
\begin{center}
\caption{Model hyper-parameters for Stacked LSTM\label{tab:table1}}
\begin{tabular}{|m{1.6cm}|m{1cm}|m{1.5cm}|m{1cm}|m{1cm}|}
\hline
\textbf{Dataset} & \textbf{No. of Layers} & \textbf{LSTM Cells/Layer} & \textbf{No. of Epochs} & \textbf{Learning Rate} \\
\hline
UNSW-NB15 & 4  & 40 128 128 \hspace{0.6em} 64 & 50 & 0.002 \\
\hline 
Bot-IoT & 2 & 32 32 & 5 & 0.002 \\
\hline 
\end{tabular}
\end{center}
\end{table}
\subsubsection{Bi-Directional LSTM} 
In case of Bi-Directional LSTMs, the recurrent network layer is replicated and it works along side the first layer. The first layer processes the input sequence, while the reversed copy of the input sequence is fed to the second layer. Learning from the past instances and the future instances provides more context to the network and results in better learning. The input layer is a Bi-Directional LSTM layer which feeds forward a non-sequential output to a Dense Layer. The Dense layer is the final layer with the number of nodes equal to the number of categories in the output label. In the decision phase the soft-max activation function is used by the dense layer to select the most probable of output classes, and the prediction error is calculated with the help of 'Sparse Categorical Crossentropy' which is then backpropogated to adjust the weights of the neural network. The model hyperparameters are shown in table 2.
\vspace{-0.3cm}
\begin{table}[H]
\begin{center}
\caption{Model hyper-parameters for Bi-LSTM\label{tab:table2}}
\begin{tabular}{| m{2cm} | m{1cm} | m{1.5cm}| m{1cm} | m{1cm}|}
\hline
\textbf{Dataset} & \textbf{No.s of Layers} & \textbf{LSTM Cells/Layer} & \textbf{No.s of Epochs} & \textbf{Learning Rate} \\
\hline
UNSW-NB15 & 1  & 64 & 50 & 0.0015 \\ 
\hline 
Bot-IoT & 1 & 12 & 5 & 0.001 \\
\hline 
\end{tabular}
\end{center}
\end{table}
\subsubsection{Model Structure and Parameters}The model is trained over a number of epochs and training and validation loss decreases gradually with time. The learning process is stopped when the number of epochs cross the maximum limit or the model starts overfitting on the training dataset.

\IEEEpubidadjcol


\section{RESULTS AND PERFORMANCE ANALYSIS}
\subsection{Experimental Setup} 
We have used the Google Colab's GPU. The specifications were Intel(R) Xeon(R) CPU with 2 cores@2.20 GHz, 12.7 GB of RAM and 78 Gb of Hard Disk space. The version of python installed was 3.7.12 and Tensorflow was 2.7.0. 
\subsection{Evaluation Metrics}
There is no single metric which can accurately tell how good a particular model is. Hence, we have used several metrics to evaluate the DL models:
\begin{itemize}
\item{
    \textbf{True Positive (TP):} Number of correctly predicted attack samples.
}
\item{
    \textbf{False Positive (FP):} Number of falsely predicted attack samples.
}
\item{
    \textbf{True Negative (TN):} Number of correctly predicted normal samples.
}
\item{
    \textbf{False Negative (FN):} Number of falsely predicted normal samples.
}
\item{
    \textbf{Accuracy:} Ratio of correctly predicted samples to total samples.
    \noindent
    \begin{align}
    ACC = \frac{TP+TN}{TP+TN+FP+FN}
    \end{align}
}
\item{
    \textbf{Precision:} Ratio of correctly predicted attack samples to total predicted attack samples.
    \noindent
    \begin{align}
    PR = \frac{TP}{TP+FP}
    \end{align}
}
\item{
    \textbf{Recall:} Ratio of correctly predicted attack samples to total number of attack samples
    \noindent
    \begin{align}
    RE = \frac{TP}{TP+FN}
    \end{align}
}
\item{
    \textbf{F1 Score:} Weighted mean of precision and recall. 
    \noindent
    \begin{align}
    F1 Score = \frac{2*Recall*Precision}{Recall+Precision}
    \end{align}
}
\item{
    \textbf{Weighted avg:} The data points which have higher frequency contribute more than others. 
}
\end{itemize}
\begin{table}[H]
\centering
\caption{Stacked LSTM\label{tab:table1}}
\begin{tabular}{ |c|c|c|c| } 
 \hline
 \textbf{Attack} & \textbf{Precision} & \textbf{Recall} & \textbf{F1} \\ 
 \hline
 Normal & 0.98 & 1.00 & 0.99 \\ 
 \hline
 Exploits & 0.56 & 0.86 & 0.67 \\ 
 \hline
 Reconnaissance & 0.79 & 0.51 & 0.62 \\ 
 \hline
 DoS & 0.87 & 0.01 & 0.01 \\ 
 \hline
 Generic & 1.00 & 0.98 & 0.99 \\ 
 \hline
 Shellcode & 0.62 & 0.41 & 0.49 \\ 
 \hline
 Fuzzers & 0.52 & 0.29 & 0.37 \\ 
 \hline
 Worms & 1.00 & 0.00 & 0.00 \\ 
 \hline
 Backdoor & 1.00 & 0.00 & 0.00 \\
 \hline
 Analysis & 1.00 & 0.00 & 0.00 \\
 \hline
 Weighted avg & 0.97 & 0.96 & 0.96 \\ 
 \hline
\end{tabular}
\end{table}
\subsection{Performance Analysis on UNSW\_NB-15 Dataset:}
The dataset was splitted into 75\% for training and 25\% for validation purpose. The training phase consisted of 50 epochs that lasted over 5 hours. In case of Stacked LSTM, the time taken for the validation phase was 128 seconds with the processing speed of 0.2 ms/sample and for Bidirectional LSTM, the validation phase took 90 seconds with the processing speed of 0.14 ms/sample.
The accuracy achieved by Stacked LSTM was 96.60\% and the accuracy achieved by Bi-directional LSTM was 96.41\%
\begin{table}[H]
\centering
\caption{Bi-directional LSTM\label{tab:table2}}
\begin{tabular}{ |c|c|c|c| } 
 \hline
 \textbf{Attack} & \textbf{Precision} & \textbf{Recall} & \textbf{F1} \\ 
 \hline
 Normal & 0.99 & 0.99 & 0.99 \\ 
 \hline
 Exploits & 0.54 & 0.79 & 0.64 \\ 
 \hline
 Reconnaissance & 0.53 & 0.59 & 0.56 \\ 
 \hline
 DoS & 0.47 & 0.03 & 0.06 \\ 
 \hline
 Generic & 1.00 & 0.98 & 0.99 \\ 
 \hline
 Shellcode & 0.68 & 0.59 & 0.64 \\ 
 \hline
 Fuzzers & 0.56 & 0.35 & 0.43 \\ 
 \hline
 Worms & 0.53 & 0.16 & 0.25 \\ 
 \hline
 Backdoor & 0.67 & 0.00 & 0.01 \\ 
 \hline
 Analysis & 1.00 & 0.00 & 0.00 \\
 \hline
 Weighted avg & 0.96 & 0.96 & 0.96 \\ 
 \hline
\end{tabular}
\end{table}
\subsection{Performance Analysis on Bot-IoT Dataset:}
In case of Bot\_IoT, the training dataset had 2,934,817 samples and testing dataset had 733,705 samples. The time taken by the training phase was 20 minutes which consisted of 5 epochs for both the stacked and bi-directional LSTM models. In case of Stacked LSTM, the validation phase lasted for 48 seconds with the processing speed of 0.06 ms/sample. For Bidirectional LSTM, the model took 156 seconds with the processing speed of 0.195 ms/sample. 
.


\begin{figure}[H]
\includegraphics[width=7cm]{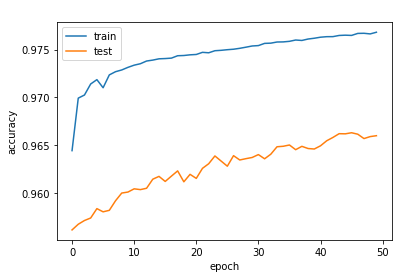}
\label{fig_first_case}
\begin{center}
\caption{Training vs.Validation accuracy (STACKED LSTM)}
\end{center}
\label{fig_sim}
\centering
\end{figure}
\vspace{-1cm}
\begin{figure}[H]
\includegraphics[width=7cm]{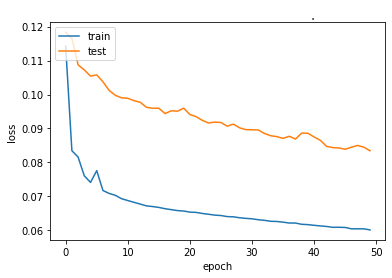}
\label{fig_first_case}
\begin{center}
\caption{Training vs. Validation loss (STACKED LSTM)}
\end{center}
\label{fig_sim}
\centering
\end{figure}

\begin{figure}[H]
\includegraphics[width=7cm]{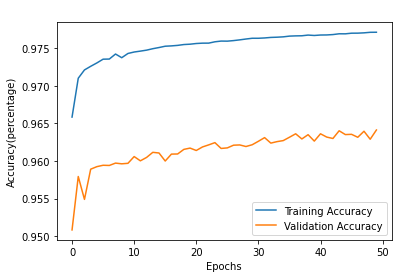}
\label{fig_first_case}
\begin{center}
\caption{Training vs. validation accuracy (BI-LSTM)}
\end{center}
\label{fig_sim}
\centering
\end{figure}
The accuracy achieved by Stacked LSTM was 99.99\% and the accuracy achieved by Bi-directional LSTM was 99.99\%.
\begin{figure}[H]
\includegraphics[width=7cm]{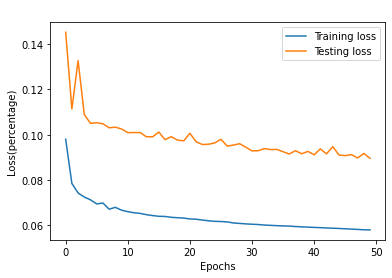}
\label{fig_first_case}
\begin{center}
\caption{Training vs.Validation loss (BI-LSTM)}
\end{center}
\label{fig_sim}
\centering
\end{figure}

\begin{table}[H]
\centering
\caption{Stacked LSTM \& Bi-directional LSTM\label{tab:table3}}
\begin{tabular}{|p{1.7cm}|p{.9cm}|p{.6cm}|p{.4cm}|p{.9cm}|p{.6cm}|p{.4cm}|}
 \hline
 \textbf{Attack} & \textbf{Precision} & \textbf{Recall} & \textbf{F1} & \textbf{Precision} & \textbf{Recall} & \textbf{F1} \\ 
 \hline
 Normal & 0.98 & 0.83 & 0.89 & 1.00 & 1.00 & 1.00 \\
 \hline
 DDoS & 1.00 & 1.00 & 1.00 & 1.00 & 1.00 & 1.00 \\ 
 \hline
 DoS & 1.00 & 1.00 & 1.00 & 1.00 & 0.79 & 0.88 \\ 
 \hline
 Reconnaissance & 1.00 & 0.93 & 0.96 & 1.00 & 0.93 & 0.96 \\ 
 \hline
 Theft & 1.00 & 0.36 & 0.53 & 1.00 & 0.36 & 0.53\\
 \hline
 Weighted avg & 1.00 & 1.00 & 1.00 & 1.00 & 1.00 & 1.00\\ 
 \hline
\end{tabular}
\end{table}

\vspace{0.3cm}

\begin{figure}[h]
\includegraphics[width=9cm]{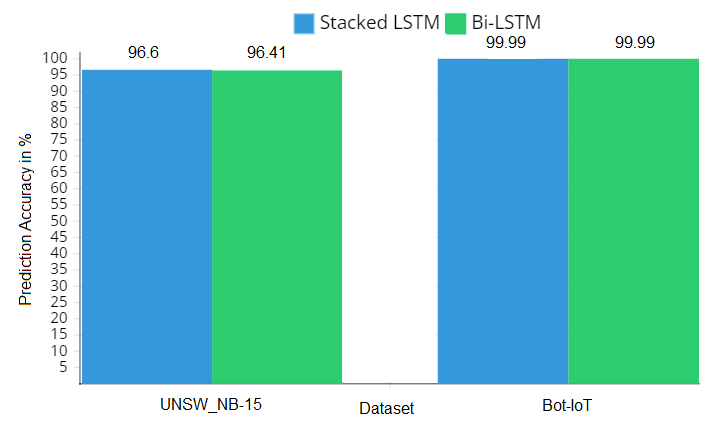}
\label{fig_first_case}
\begin{center}
\caption{NIDS accuracy comparision of Bi-LSTM and Stacked LSTM on both datasets}
\end{center}
\label{fig_sim}
\centering
\end{figure}
\vspace{-0.3cm}
\subsection{Comparison with other techniques:}

\begin{table}[H]
\centering
\caption{Comparison of Results on BoT-IoT and UNSW\_NB-15  Datasets\label{tab:table4}}
\begin{tabular}{|c|c|c|c|}
\hline
\multicolumn{2}{|c} {BoT-IoT} & \multicolumn{2}{|c|} {UNSW\_NB-15}\\
\hline
\textbf{Method} & \textbf{Accuracy} (\%) & \textbf{Method} & \textbf{Accuracy} (\%) \\ 
\hline
 ARM \hyperlink{ref5}{[5]} & 85.6 & FSVM \hyperlink{ref4}{[4]} & 92 \\
 \hline
 Decision Tree \hyperlink{ref5}{[5]} & 93.2 & GAA \hyperlink{ref4}{[4]} & 93\\ 
 \hline
 DNN \hyperlink{ref5}{[5]} & 99.9 & DNN \hyperlink{ref5}{[5]} & 99.2\\ 
 \hline
 Naive Bayes \hyperlink{ref5}{[5]} & 72.7 & ANN \hyperlink{ref6}{[6]} & 63.97\\
 \hline
 Perceptron \hyperlink{ref5}{[5]} & 63.9 & ARM \hyperlink{ref6}{[6]} & 86.45 \\
 \hline
 - & - & RNN \hyperlink{ref8}{[8]} & 95.7 \\
 \hline
 - & - & Hybrid \hyperlink{ref8}{[8]} & 98.7 \\
 \hline
 {\bf LBDMIDS} & {\bf 99.9} & {\bf LBDMIDS} & {\bf 96.6} \\ 
 \hline
\end{tabular}
\end{table}
The results in paper \hyperlink{ref4}{[4]}, \hyperlink{ref6}{[6]} and \hyperlink{ref8}{[8]} are for binary classification (Attack and Benign) while the results in paper \hyperlink{ref5}{[5]} and proposed methodology (LBDMIDS) is for Multi-Class classification.

 
%
\section{Conclusion and Future Works}
 To protect the IoT networks from  attackers and their attacks, it is necessary to detect the intrusions precisely.    
In this paper, a DL method based model called LBDMIDS has been proposed which shows promising performance in intrusion detection.

\par 
In this paper, the focus was on detection of malicious events with improved accuracy in IoT networks where the dataset is large and time series. LBDMIDS works on LSTM architecture to detect intrusion in a network. To validate LBDMIDS, the experiment was performed on two well known datasets, i.e., BoT-IoT and UNSW-NB15. We scaled and normalized the dataset accordingly and fed it to LBDMIDS. The output and results produced by LBDMIDS are good in terms of prediction accuracy and F1-score. To generalize our model in LBDMIDS, Stacked and Bidirectional LSTM were used on UNSW-NB15 dataset, the accuracy achieved by Stacked LSTM model was 96.60\% and the accuracy achieved by Bi-Directional LSTM model was 96.41\%. Similarly, on BoT-IoT dataset, the accuracy achieved by Stacked and Bi-Directional LSTM model was 99.99\%.
\par Owing to the system limitations and epoch duration, the accuracy could be significantly improved considering more efficient and robust machines. Our experiment was not performed on industrial scale or real world IoT applications \hyperlink{ref20}{[20]}. With more efficient GPUs and more hybrid DL models, the prediction accuracy could be improved.
 


 




\vfill

\end{document}